\documentclass[journal]{IEEEtran}

\usepackage{cite}
\usepackage{amsmath,amssymb}
\usepackage{graphicx}
\usepackage{booktabs}
\usepackage{url}
\usepackage[hidelinks]{hyperref}

\begin{document}

\title{Cyclic-Prefix-Free OFDM With Tail-Reuse Reconstruction for Distributed Acoustic Sensing}

\author{Ziang Chen, Zhiyang Xue, Zhongxing Tian, Zeyu Feng, Dongdong Zou,\\ 
          Yi Cai*, \emph{Fellow, Optica} and Huan Huang*, \emph{Member, IEEE}%
\thanks{Manuscript received Month Day, Year; revised Month Day, Year. (Corresponding authors:
Yi Cai; Huan Huang.)}
\thanks{The authors are with Soochow University, Suzhou, China (e-mail:
\mbox{yicai@suda.edu.cn}; \mbox{huanghuan@suda.edu.cn}).}
}

\markboth{Journal of Lightwave Technology,~Vol.~XX, No.~XX, Month~2026}%
{Chen \MakeLowercase{\textit{et al.}}: Cyclic-Prefix-Free OFDM With Tail-Reuse Reconstruction for DAS}

\maketitle

\begin{abstract}
Orthogonal frequency-division multiplexing (OFDM) enables frequency-domain reconstruction of the
distributed Rayleigh backscatter channel in coherent distributed acoustic sensing (DAS). However,
an explicitly transmitted cyclic prefix (CP) lengthens the probing period and lowers the slow-time
Nyquist limit of each range cell. Here, we investigate a repeated cyclic-prefix-free OFDM (No-CP
OFDM) waveform for DAS, using the preceding useful-period tail as a virtual cyclic extension. We
derive the finite-memory range condition for tail-reuse reconstruction and identify circular folding
when the useful period is shorter than the physical channel memory. Analysis shows that, for a fixed useful-block length
and fiber memory, removing the explicit CP raises the period-limited highest unaliased vibration
frequency without altering the occupied-bandwidth-limited spatial resolution. For a 5.2-km numerical configuration, a 75-MSa/s
processing rate and a 4096-sample useful block yield a 54.61-$\mu$s No-CP period and a 9.16-kHz
slow-time Nyquist limit. Simulations verify tail reuse and the predicted range-folding boundary, and demonstrate recovery
of 100 vibration events spanning 800~Hz to 8800~Hz. A bandwidth-scaling simulation further shows
that jointly processing fine spatial observations improves differential-phase reliability and
reconstruction SNR at a fixed reporting interval under the modeled conditions. Experiments on a 5.2-km coherent DAS link with
111.984-MHz occupied OFDM
bandwidth blindly localize 500-Hz and 3-kHz piezoelectric transducer (PZT) vibrations at 5063.7~m
and 5070.1~m, respectively. The corresponding waveforms and spectra are recovered from the detected
range bins. These experiments establish the feasibility of tail-reuse channel reconstruction and
vibration recovery, while the period-limited extension of the unaliased vibration bandwidth is
quantified analytically and verified numerically.
\end{abstract}


\IEEEpeerreviewmaketitle

\section{Introduction}
\IEEEPARstart{D}{istributed} acoustic sensing (DAS) based on coherent Rayleigh backscatter enables
optical fiber to serve not only as a transmission medium but also as a distributed array of
vibration-sensitive sensing points over large numbers of range bins. In a phase-sensitive optical
time-domain reflectometry ($\phi$-OTDR) system, an optical probe
is launched into the fiber, the weak Rayleigh backscatter is coherently detected, and dynamic strain
or vibration is inferred from the slow-time evolution of the complex backscattered field at each
range position \cite{posey2000strain,lu2010phaseotdr}. Because the sensing element is the fiber
itself, DAS can use deployed or purpose-installed cables to monitor long linear assets and extended
physical environments without placing individual electrical sensors along the path. This capability
has motivated distributed fiber-optic sensing for infrastructure, perimeter, geophysical, and
seismic monitoring applications, including demonstrations using existing dark fiber for near-surface
characterization and broadband event detection
\cite{bao2012distributed,masoudi2016dynamicstrain,muanenda2018recent,ajofranklin2019darkfiber,he2021dasreview}.

The usefulness of DAS is determined not only by whether a vibration can be detected, but also by
how sensing range, spatial resolution, signal-to-noise ratio (SNR), and vibration bandwidth are
obtained simultaneously. These requirements are coupled. A longer fiber increases propagation
delay, attenuation, and the temporal memory of the backscatter channel; a finer spatial resolution
requires a shorter effective impulse response or larger waveform bandwidth; a higher SNR usually
requires more optical energy, processing gain, or coherent averaging; and a higher vibration
frequency requires faster slow-time sampling of each range cell. Therefore, a DAS waveform is not
merely an illumination signal. It also defines the range-domain point-spread function, the amount
of useful energy per observation, the repetition period available for dynamic sampling, and the
receiver operation used to convert backscatter into a spatial trace. This coupling makes the
probing waveform and the reconstruction receiver central design choices in long-reach,
high-bandwidth DAS.

Conventional short-pulse DAS offers a direct and intuitive range mapping: the pulse duration sets
the nominal spatial resolution, while the round-trip delay maps each backscatter sample to a fiber
position. This simplicity is one reason that pulse-based $\phi$-OTDR remains a widely used
architecture for distributed vibration sensing. However, the same mechanism also produces a basic
energy--resolution tension. Increasing pulse width can inject more optical energy and improve SNR,
but it broadens the spatial response; shortening the pulse improves spatial resolution, but reduces
the available pulse energy unless the peak power is increased. In long-distance or high-bandwidth
DAS, this tradeoff becomes more severe because the system must maintain enough backscattered SNR
while preserving spatial selectivity and refreshing each range cell fast enough for dynamic
measurements \cite{bao2012distributed,masoudi2016dynamicstrain}.

Pulse-compression and coding techniques have been introduced to relax this tension by increasing
the time--bandwidth product or coding gain without simply broadening the final compressed response.
Early optical time-domain reflectometry studies used pulse-compression waveforms to improve the
range-resolved measurement process, and related ideas have been extended to phase-sensitive OTDR
for sub-meter-resolution vibration sensing
\cite{yang2014pulsecompressionotdr,lu2017pulsecompressionphiotdr}. Chirped-pulse and swept-wave
approaches further exploit frequency modulation to encode range-dependent phase or strain
information, while coded-pulse schemes use structured sequences to improve sensitivity and reach
\cite{gonzalez2016chirpedpulse,pastor2016chirped,fernandez2019chirped,wang2019pulsecoding,li2021pulsecoding,jiang2021continuouschirped}.
These developments show that waveform design is an effective way to redistribute optical energy,
bandwidth, and processing gain in DAS.

Despite these advantages, many pulse-compression, chirped, and coded DAS receivers still recover
the range trace through linear correlation with a known replica of the transmitted waveform. The
recovered trace is therefore the fiber channel convolved with the aperiodic waveform
autocorrelation. A narrow main lobe supports fine nominal spatial resolution, whereas sidelobes
and finite main-lobe width spread the response of one fiber section into neighboring sections. In a
dense Rayleigh backscatter channel, these off-peak correlation terms produce deterministic
inter-range leakage that remains after coherent averaging. This motivates a receiver model that
treats the backscattered fiber as a finite-memory sensing channel and makes the effective range
response explicit in the reconstruction process.

Frequency-domain channel reconstruction provides a different viewpoint. In discrete
Fourier-transform OFDM, the transmitted block is synthesized from orthogonal subcarrier weights,
and a cyclic prefix (CP) is inserted so that the useful block sees the finite-memory channel as a
circular convolution. After the discrete Fourier transform, the received sample on each active
subcarrier is proportional to the corresponding channel frequency response. This converts a
time-domain convolution problem into subcarrier-wise channel estimation or equalization, a principle
that has been central to multicarrier communication systems and later optical OFDM implementations
\cite{weinstein1971dftofdm,cimini1985mobileofdm,bingham1990multicarrier,shieh2006coherentofdm,lowery2006ofdmoptical,armstrong2009ofdm}.

When this principle is transferred from communications to DAS, the estimated channel takes on a
different role. In data transmission, the channel is an impairment to be equalized; in DAS, the
range-domain Rayleigh response is itself the sensing quantity. Our preceding CP-OFDM DAS study used
known OFDM subcarriers to estimate the active-bin channel response and reconstructed the range-domain
channel by inverse Fourier transform \cite{huang2026cpofdmisi}. The effective range response is
therefore governed by subcarrier inversion and the active-frequency aperture rather than only by the
aperiodic probe autocorrelation. This reconstruction framework also exposes a DAS-specific cost: the
CP lengthens each slow-time frame without providing an additional useful OFDM observation.

In DAS, the CP therefore imposes a sensing penalty. The OFDM block repetition period is the
slow-time sampling interval of every recovered range cell. At a fixed sampling rate and useful-block
length, adding a CP lengthens each channel observation and lowers its slow-time Nyquist limit. Because
the prefix must cover the effective fiber memory, it can occupy a substantial fraction of the probing
period over a long fiber. The same prefix that establishes circular convolution for frequency-domain
reconstruction therefore reduces the vibration bandwidth available to each range cell.

This penalty is relevant to broadband, high-frequency, and high-spatial-resolution DAS
\cite{fernandez2019chirped,marcon2019ofdrdas,jiang2021continuouschirped,he2021dasreview}. Optical
frequency-domain, swept/chirped, and pulse-coded approaches address these objectives through
different waveform and receiver tradeoffs. Within the OFDM channel-reconstruction framework,
however, it remains unclear whether the required circular-convolution condition can be obtained
without an explicitly transmitted CP. Resolving this issue would retain frequency-domain channel
reconstruction while avoiding prefix samples that lower the highest unaliased vibration frequency.

Here, we investigate repeated cyclic-prefix-free OFDM (No-CP OFDM) with tail-reuse reconstruction
for DAS. The known useful period is transmitted continuously so that the preceding-period tail
provides the virtual cyclic extension for the current period. The receiver then estimates the
active-bin channel response and reconstructs the range-domain channel without a separately
transmitted prefix. Tail
reuse therefore retains the CP-OFDM channel-reconstruction interpretation while shortening the
slow-time observation period.

The main contributions are as follows. First, we formulate a finite-memory model for repeated No-CP
OFDM DAS and identify the condition under which the received useful block follows the circular-convolution
relation required for frequency-domain channel reconstruction. Second, we connect explicit-CP
overhead to slow-time vibration bandwidth and range folding, thereby identifying when useful-period
repetition supplies the required cyclic extension. We also distinguish delay-grid sampling from
bandwidth-limited spatial resolution. Third, simulations verify the predicted range boundary and
high-frequency vibration recovery. They further evaluate the improvement in differential-phase
reliability and reconstruction SNR obtained by jointly processing fine spatial observations enabled
by bandwidth scaling at a fixed reporting interval. Finally, we implement a tail-reuse No-CP OFDM DAS
receiver and experimentally demonstrate blind localization and frequency recovery for 500-Hz and
3-kHz vibrations over a 5.2-km fiber link.

\section{Principle and Signal Model}
\label{sec:principle}

\begin{figure}[!h]
\centering
\includegraphics[width=\columnwidth]{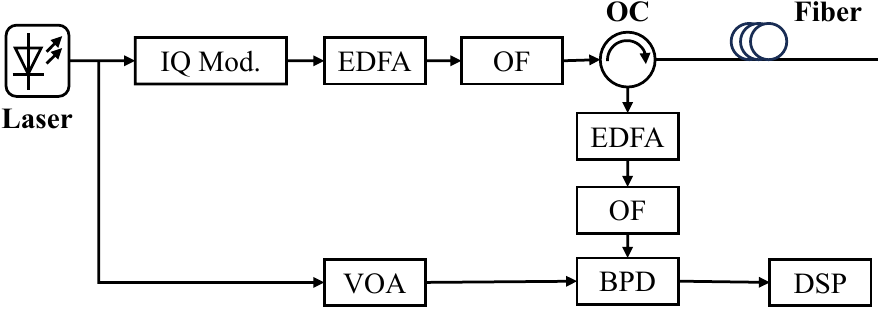}
\caption{A coherent DAS system used for repeated cyclic-prefix-free OFDM (No-CP OFDM) probing.
The inset shows three repeated No-CP OFDM useful blocks without inserted CP samples. IQ: in-phase/quadrature; 
EDFA: erbium-doped fiber amplifier; OF: optical filter; OC: optical circulator; 
BPD: balanced photodiode; DSP: digital signal processing.}
\label{fig:system-schematic}
\end{figure}

Fig.~\ref{fig:system-schematic} summarizes the coherent $\phi$-OTDR DAS architecture considered in
this work. The section begins with a finite-memory sensing-channel model for a generic probing
sequence and shows how correlation readout determines the effective range response. It then
specializes the probe to repeated No-CP OFDM and derives the tail-reuse circular-convolution
condition. The resulting design discussion connects the range-validity condition to CP overhead,
slow-time vibration bandwidth, and unambiguous range, while distinguishing delay-grid sampling from
bandwidth-limited spatial resolution.

\subsection{Finite-Memory DAS Channel and Effective Range Response}
\label{subsec:finite-memory-channel}
The starting point is the finite-memory sensing-channel view used in the CP-OFDM DAS precursor
\cite{huang2026cpofdmisi}. Over one slow-time frame $p$, the coherent Rayleigh backscatter from the
fiber is represented by the discrete range-domain channel
$\mathbf{h}_p=[h_p[0],h_p[1],\ldots,h_p[M-1]]^{\mathrm{T}}$, where $M$ is the number of spatial cells
covered by the fiber under test. With sampling interval $T_s$ and group velocity $v_g$, one discrete
delay bin corresponds to the two-way spatial increment $\Delta z=v_gT_s/2$. For a transmitted
baseband probing sequence $x[n]$, the received sequence in frame $p$ is modeled as
\begin{equation}
  y_p[n] = \sum_{m=0}^{M-1} h_p[m] x[n-m] + w_p[n],
  \label{eq:linear-channel}
\end{equation}
where $w_p[n]$ includes receiver noise and unmodeled impairments. The vibration-induced phase
variation is embedded in the slow-time evolution of $h_p[m]$.

This model also clarifies the effective range response of conventional correlation-based DAS. If a
finite-duration probing waveform $s[n]$ is recovered through linear correlation with its known
replica, let $r_s[\ell]$ denote its aperiodic autocorrelation, $E_s=r_s[0]$ its energy, and
$c_s[\ell]=r_s[\ell]/E_s$ its normalized autocorrelation response. The energy-normalized
matched-filter estimate at range bin $\ell$ can then be written as
\begin{equation}
\begin{aligned}
  \hat{h}^{\mathrm{MF}}_p[\ell]
  &= \sum_{m=0}^{M-1} h_p[m]c_s[\ell-m]+\tilde{\nu}_p[\ell]\\
  &= h_p[\ell]
  +\sum_{\substack{m=0\\m\ne \ell}}^{M-1}h_p[m]c_s[\ell-m]
  +\tilde{\nu}_p[\ell].
\end{aligned}
  \label{eq:mf-spatial-isi}
\end{equation}
The summation term in \eqref{eq:mf-spatial-isi} is the waveform-dependent leakage term: range bins
other than $\ell$ enter the recovered bin through the off-peak samples of $c_s[\ell]$. A finer
analog-to-digital converter (ADC) grid may represent this point-spread function with more samples,
but it does not change the underlying autocorrelation-weighted response. The prior CP-OFDM
framework instead forms a circular-convolution observation, estimates the loaded frequency bins,
and transforms them back to the range domain \cite{huang2026cpofdmisi}. Its effective range kernel
is determined by the subcarrier inversion and active-frequency weighting. The present repeated
No-CP work keeps this reconstruction view but replaces the explicitly transmitted cyclic prefix
with useful-symbol repetition.

\subsection{Repeated No-CP OFDM Tail-Reuse Reconstruction}
\label{subsec:periodic-nocp}
Let $N$ denote the OFDM useful-symbol length and let $S[k]$ be the known complex weight on the
$k$th subcarrier. One useful OFDM period is
\begin{equation}
  s[n] = \frac{1}{N}\sum_{k=0}^{N-1} S[k] e^{j2\pi kn/N},
  \quad 0 \le n < N.
  \label{eq:ofdm-useful}
\end{equation}
In the proposed repeated No-CP sensing waveform, no cyclic prefix, cyclic suffix, or zero guard is inserted.
Instead, the same known useful OFDM period is transmitted continuously,
\begin{equation}
  x[n] = s[\langle n\rangle_N],
  \quad x[n+N]=x[n],
  \label{eq:periodic-transmission}
\end{equation}
where $\langle n\rangle_N$ denotes the integer $n$ modulo $N$.

\begin{figure}[!h]
\centering
\includegraphics[width=\columnwidth]{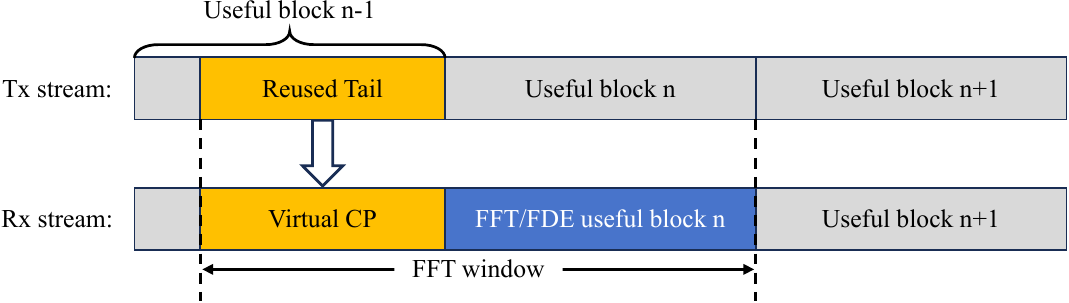}
\caption{Principle of virtual-CP construction in repeated No-CP OFDM with tail reuse. The
transmitted stream contains only repeated useful OFDM blocks. The tail of useful block $n-1$ is reused at the receiver
as the virtual CP for useful block $n$, after which the useful block is processed by FFT/FDE. FFT: fast Fourier
transform; FDE: frequency-domain equalization.}
\label{fig:tail-reuse-principle}
\end{figure}

The corresponding timing relation is illustrated in Fig.~\ref{fig:tail-reuse-principle}. Unlike
conventional CP-OFDM, the highlighted tail is not a newly inserted guard interval; it is already part
of the preceding useful OFDM period in the continuous transmitted stream. At the receiver, this tail
is used as the virtual cyclic extension associated with the current useful period before channel
reconstruction based on the fast Fourier transform (FFT) and frequency-domain equalization (FDE).

The key observation formalized below is that a repeated useful period provides a virtual cyclic
extension. For the current period, the last $M-1$ samples of the previous period are identical to
the samples that would have been copied into an explicit cyclic prefix. Thus no additional prefix
samples are transmitted, but the linear fiber convolution can still be observed as a circular
convolution when the repeated period covers the channel memory. Under the range condition
\begin{equation}
  N \ge M,
  \label{eq:period-length-condition}
\end{equation}
and assuming that the channel is approximately constant over the previous-tail-plus-current-useful
observation interval, an aligned $N$-sample receive window becomes
\begin{equation}
  u_p[n] = \sum_{m=0}^{M-1} h_p[m] s[\langle n-m\rangle_N] + w_p[n],
  \quad 0 \le n < N.
  \label{eq:circular-channel}
\end{equation}
Equation~\eqref{eq:circular-channel} is the theoretical bridge from sufficient-CP OFDM to repeated
No-CP OFDM: the circular-convolution condition is preserved by repeating the known useful period,
while the explicit CP is removed from the transmitted stream.

Taking the $N$-point discrete Fourier transform (DFT) of \eqref{eq:circular-channel} gives
\begin{equation}
\begin{aligned}
  U_p[k] &= S[k]H_p[k]+W_p[k],\\
  H_p[k] &= \sum_{m=0}^{M-1}h_p[m]e^{-j2\pi km/N}.
\end{aligned}
  \label{eq:freq-domain-channel}
\end{equation}
For active subcarriers $k\in\mathcal{K}$, the channel spectrum is estimated by one-tap
regularized least squares (LS),
\begin{equation}
  \hat{H}_p[k] =
  \begin{cases}
    \dfrac{U_p[k]S^{*}[k]}{|S[k]|^2+\lambda}, & k\in\mathcal{K},\\[2mm]
    0, & k\notin\mathcal{K},
  \end{cases}
  \label{eq:regularized-fde}
\end{equation}
where $\lambda$ controls noise enhancement near weak loaded subcarriers. The case $\lambda=0$
corresponds to zero-forcing (ZF) inversion on the active set. The range-domain channel estimate is
then obtained by inverse DFT,
\begin{equation}
  \hat{h}_p[m] = \frac{1}{N}\sum_{k=0}^{N-1}\hat{H}_p[k]e^{j2\pi km/N},
  \quad 0 \le m < M.
  \label{eq:idft-channel}
\end{equation}

The effective range response follows directly from
\eqref{eq:freq-domain-channel}--\eqref{eq:idft-channel}. Define the deterministic frequency-domain
weighting as $G_{\lambda}[k]$. Then
\begin{equation}
\begin{aligned}
  G_{\lambda}[k]
  &=
  \begin{cases}
    \dfrac{|S[k]|^2}{|S[k]|^2+\lambda}, & k\in\mathcal{K},\\[1mm]
    0, & k\notin\mathcal{K},
  \end{cases}\\
  g_{\lambda}[m]
  &= \frac{1}{N}\sum_{k=0}^{N-1}
  G_{\lambda}[k]
  e^{j2\pi km/N},\\
  \hat{h}_p[m]
  &= \sum_{q=0}^{M-1} h_p[q]
  g_{\lambda}[\langle m-q\rangle_N]
  + \tilde{w}_{p,\lambda}[m].
\end{aligned}
\label{eq:effective-reconstruction-kernel}
\end{equation}
Here $g_{\lambda}[m]$ is the circular point-spread function, and
$\tilde{w}_{p,\lambda}[m]$ is the correspondingly weighted noise. Equation~\eqref{eq:effective-reconstruction-kernel}
provides a common description of full-bin reconstruction, inactive subcarriers, and regularization.

\emph{Proposition 1 (tail reuse for repeated No-CP reconstruction):} Suppose that the same useful
OFDM period is repeated, $N\ge M$, the receiver window is aligned to a consistent periodic windowing
phase, and the channel is quasi-static over the effective previous-tail-plus-current-useful
observation interval. In the ideal full-bin case with $S[k]\ne 0$ for all $k$ and $\lambda=0$, the
ZF reconstruction in \eqref{eq:idft-channel} gives
\begin{equation}
  \hat{h}_p[m] = h_p[m] + \tilde{w}_p[m],
  \quad 0 \le m < M,
  \label{eq:nocp-proposition}
\end{equation}
where $\tilde{w}_p[m]$ represents receiver noise after frequency-domain inversion and IDFT. If the noise term is absent,
$\hat{h}_p[m]=h_p[m]$ on the DFT delay grid, and no deterministic contribution from $q\ne m$ range
bins remains.

To see this, we substitute the definition of $H_p[k]$ in \eqref{eq:freq-domain-channel} into
\eqref{eq:idft-channel} with $\lambda=0$ and full nonzero subcarrier loading, and obtain
\begin{equation}
  \hat{h}_p[m]
  =
  \sum_{q=0}^{M-1} h_p[q]
  \left[
  \frac{1}{N}\sum_{k=0}^{N-1}e^{j2\pi k(m-q)/N}
  \right]
  +
  \tilde{w}_p[m].
  \label{eq:nocp-proof-sum}
\end{equation}
The bracketed sum evaluates to $\delta_N[m-q]$, the Kronecker delta on an $N$-periodic delay grid.
On the physical fiber support, the condition $N\ge M$ prevents modular wraparound between two
different range bins, so $\delta_N[m-q]$ acts as an ordinary Kronecker delta. The first term in
\eqref{eq:nocp-proof-sum} is therefore $h_p[m]$. The key distinction from explicit-CP OFDM is not
the frequency-domain estimator itself, but the origin of the cyclic extension. Here it is supplied
by tail reuse of the preceding repeated period rather than by newly transmitted prefix samples. For
constant-modulus full-bin loading, the zero-regularization estimator in
\eqref{eq:regularized-fde} is, up to a constant scale, mathematically equivalent to circular matched
filtering. Exact discrete-channel recovery then follows because the cyclic autocorrelation is a
Kronecker delta. With inactive bins or regularization, the effective response is instead
$g_{\lambda}[m]$ in \eqref{eq:effective-reconstruction-kernel}.

The DAS vibration readout is obtained from the reconstructed channel through gauge-differential
phase processing. For a gauge separation of $G$ samples, the complex gauge product and its phase
change relative to a reference frame $p=0$ are
\begin{equation}
\begin{aligned}
  \hat{g}_p[m] &= \hat{h}_p[m+G]\hat{h}_p^{*}[m],
  \quad 0 \le m < M-G,\\
  \Delta\phi_p[m] &= \angle\left\{\hat{g}_p[m]\hat{g}_0^{*}[m]\right\}.
\end{aligned}
  \label{eq:gauge-phase}
\end{equation}
This operation converts slow-time changes in the reconstructed Rayleigh channel into local vibration
traces along the fiber.

\subsection{Design Consequences: Overhead, Vibration Bandwidth, and Range}
\label{subsec:design-consequences}
The repeated No-CP model links cyclic-extension overhead, slow-time sampling, and range validity in
DAS operation. For a conventional CP-OFDM sensing waveform using the same useful-symbol length $N$,
a sufficient CP must cover the finite channel memory,
\begin{equation}
  L_{\mathrm{CP}} \ge M-1.
  \label{eq:sufficient-cp-length}
\end{equation}
The corresponding CP-OFDM and No-CP repetition periods are
\begin{equation}
  T_{\mathrm{CP}} = (N+L_{\mathrm{CP}})T_s,
  \quad
  T_{\mathrm{NoCP}} = NT_s.
  \label{eq:period-comparison}
\end{equation}
Thus the useful-symbol-normalized CP overhead is $L_{\mathrm{CP}}/N$, and the fraction of a
sufficient-CP frame occupied by CP samples is $L_{\mathrm{CP}}/(N+L_{\mathrm{CP}})$. The No-CP
waveform does not remove the physical requirement that the channel memory be covered; instead, it
reuses the previous repeated useful-period tail as the virtual cyclic extension.

The second consequence is an increase in the slow-time sampling rate relative to an explicit-CP
implementation with the same $N$ and fiber memory. Since DAS vibration recovery samples the channel
once per repeated probing period, the highest unaliased vibration frequency is bounded by
\begin{equation}
  f_{\max} = \frac{1}{2T_{\mathrm{rep}}}.
  \label{eq:vibration-nyquist}
\end{equation}
Therefore,
\begin{equation}
  \frac{f_{\max}^{\mathrm{NoCP}}}{f_{\max}^{\mathrm{CP}}}
  =
  \frac{T_{\mathrm{CP}}}{T_{\mathrm{NoCP}}}
  =
  1+\frac{L_{\mathrm{CP}}}{N}.
  \label{eq:vibration-bandwidth-gain}
\end{equation}
The range-validity condition further links fiber length to slow-time bandwidth. Because
$T_{\mathrm{NoCP}}=NT_s$ and $N\geq M$,
\begin{equation}
  T_{\mathrm{NoCP}}
  \geq MT_s
  \approx \frac{2L}{v_g},
  \quad
  f_{\max}^{\mathrm{NoCP}}
  \lesssim \frac{v_g}{4L},
  \label{eq:length-limited-vibration-bandwidth}
\end{equation}
where $L$ is the fiber length. The round-trip propagation time therefore sets a lower bound on the
valid useful-period duration and an upper bound on the unaliased vibration frequency. By eliminating
a separately transmitted prefix, tail reuse allows repeated No-CP OFDM to approach this
length-constrained limit. Nevertheless, the useful period must still span the full fiber round-trip
memory to avoid range folding.

For a minimum-sufficient explicit-CP reference, substituting
$L_{\mathrm{CP}}\ge M-1$ into \eqref{eq:vibration-bandwidth-gain} gives the
period penalty needed to accommodate the same fiber memory. For completeness, if a CP-OFDM sensing
implementation additionally inserts a guard or idle interval of $L_{\mathrm{g}}$ samples between
probing blocks, its repetition period and the corresponding vibration-bandwidth ratio become
\begin{equation}
  T_{\mathrm{CP,rep}}=(N+L_{\mathrm{CP}}+L_{\mathrm{g}})T_s,
  \quad
  \frac{f_{\max}^{\mathrm{NoCP}}}{f_{\max}^{\mathrm{CP,rep}}}
  =
  \frac{N+L_{\mathrm{CP}}+L_{\mathrm{g}}}{N}.
  \label{eq:cp-guard-comparison}
\end{equation}
Equation~\eqref{eq:cp-guard-comparison} generalizes the period comparison; the continuous CP-OFDM
reference used in Section~\ref{sec:numerical-simulation} corresponds to $L_{\mathrm{g}}=0$.

The same design variables also set the unambiguous circular-deconvolution range,
\begin{equation}
  \Delta z=\frac{v_gT_s}{2}=\frac{v_g}{2F_s},
  \quad
  R_{\mathrm{amb}} = N\Delta z.
  \label{eq:unambiguous-range}
\end{equation}
Here $\Delta z$ is the DFT delay-grid sampling interval. The useful-block length must be
chosen so that $R_{\mathrm{amb}}$ exceeds the physical fiber support; any excess unambiguous range is a
range-validity margin rather than a spatial-resolution gain. The physical resolving interval is instead
set by the occupied frequency aperture,
\begin{equation}
  B_{\mathrm{occ}} = N_{\mathrm{act}}\Delta f,
  \quad
  \Delta z_{\mathrm{res}}\approx \frac{v_g}{2B_{\mathrm{occ}}},
  \label{eq:bandwidth-limited-resolution}
\end{equation}
where $N_{\mathrm{act}}$ is the number of active subcarriers and $\Delta f$ is the subcarrier
spacing. Thus the DFT delay-grid sampling interval, the unambiguous periodic range, and the
bandwidth-limited resolving interval are related but distinct quantities.

Bandwidth expansion must be considered jointly with the processing rate and useful-block length. A
wider probing band generally requires a higher complex-baseband processing rate $F_s=1/T_s$, and the
resulting finer delay grid increases the discrete channel memory according to
\begin{equation}
  M \approx \frac{2LF_s}{v_g}.
  \label{eq:bandwidth-memory-scaling}
\end{equation}
The useful-block length must therefore increase with $F_s$ to preserve
$N/F_s\geq 2L/v_g$. Scaling $N$ and $F_s$ proportionally leaves the useful-period duration,
unambiguous sensing range, slow-time sampling rate, and vibration Nyquist limit unchanged. Because
the minimum valid useful-period duration is constrained by the fiber round-trip memory, this
proportional scaling does not provide a further increase in the slow-time Nyquist limit. The
additional occupied bandwidth is instead expressed as a finer intrinsic spatial resolving interval.

If the output interval is deliberately held coarser than this intrinsic resolution, a reported
spatial unit contains multiple resolvable fine-grid observations. Their nominal ratio is
\begin{equation}
  \rho_B =
  \frac{B_{\mathrm{occ}}}{B_{\mathrm{tar}}},
  \quad
  B_{\mathrm{tar}}\approx\frac{v_g}{2\Delta z_{\mathrm{tar}}},
  \label{eq:bandwidth-redundancy}
\end{equation}
where $\Delta z_{\mathrm{tar}}$ is the target effective spatial resolution and $B_{\mathrm{tar}}$
is the bandwidth required to support that resolution. Here, $\rho_B$ characterizes the available
spatial sampling ratio, not the number of statistically independent observations, because adjacent
gauge-phase traces can remain correlated. Fine-grid traces with locally consistent slow-time
responses may be gain-aligned and combined within each reporting cell without temporal averaging,
which can reduce sensitivity to individual Rayleigh fading nulls while preserving the frame rate.
The gain depends on power normalization, receiver noise bandwidth, fading statistics, and
inter-trace correlation. This processing opportunity is not unique to repeated No-CP OFDM; tail reuse
allows it to be retained without adding CP-induced slow-time overhead.

\subsection{Validity Boundary and Practical Interpretation}
\label{subsec:validity-boundary}
The repeated No-CP principle requires the same known OFDM useful period to be transmitted
continuously. If the useful period
changes from frame to frame, the previous period tail is no longer the cyclic continuation of the
current period and \eqref{eq:circular-channel} no longer follows. The channel must also be
quasi-static over the effective previous-tail-plus-useful observation interval.

The period-length condition in \eqref{eq:period-length-condition} defines the most important failure
mode. If $N<M$, the receiver cannot uniquely assign all physical ranges to one DFT delay grid. The
periodic deconvolution then recovers a folded channel,
\begin{equation}
  h_{p,\mathrm{fold}}[r]
  =
  \sum_{\ell:\, r+\ell N < M} h_p[r+\ell N],
  \quad 0 \le r < N,
  \label{eq:folded-channel}
\end{equation}
so scatterers or perturbations separated by integer multiples of $N\Delta z$ contribute to the same
reconstructed bin. This folding is a range ambiguity of the periodic probing waveform, not a
receiver implementation artifact.

Even when $N\ge M$, the receiver must select a periodic windowing phase that maps the circularly
reconstructed channel to the physical fiber span. When $N>M$ and the expected physical support is
known, a useful timing metric is the ratio between reconstructed energy inside that support and
residual energy in the circular tail,
\begin{equation}
  \eta(q) =
  \frac{\sum_{m=0}^{M-1}|\hat{h}_{p,q}[m]|^2}
       {\sum_{m=M}^{N-1}|\hat{h}_{p,q}[m]|^2+\epsilon},
  \label{eq:tail-metric}
\end{equation}
where $q$ denotes a candidate window start and $\epsilon$ prevents division by zero. Maximizing
$\eta(q)$ favors the alignment in which the reconstructed channel energy lies within the physical
fiber span rather than wrapping into the unused DFT tail.

With edge guards, DC nulling, transmitter and receiver frequency responses, active-bin
regularization, and finite SNR, the practical reconstruction is a band-limited and regularized
channel estimate. A grid-aligned point scatterer can be reconstructed as a Kronecker-delta response
on the DFT delay grid under Proposition 1, but an off-grid scatterer or a
reduced active-frequency aperture produces a finite-bandwidth Dirichlet- or windowed-sinc-like
response whose first-null scale is governed by $1/B_{\mathrm{occ}}$. The proposed repeated No-CP scheme
uses periodic circular deconvolution to remove explicit CP overhead and increase the slow-time
sampling rate under the stated range and timing conditions, while the spatial resolution remains
governed by the occupied bandwidth.

\begin{figure}[!t]
\centering
\includegraphics[width=\columnwidth]{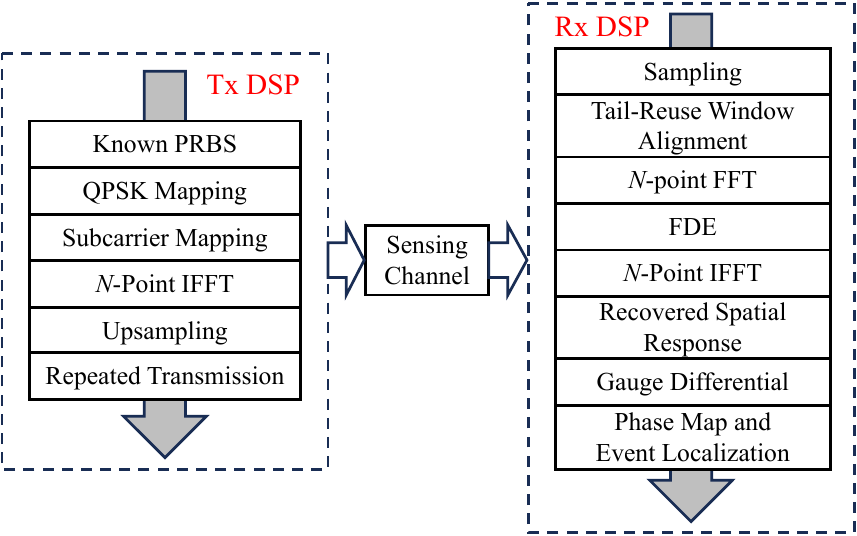}
\caption{Transmitter and receiver DSP flow used in the repeated No-CP OFDM DAS simulation, from waveform
generation to gauge-differential phase recovery. PRBS: pseudo-random binary sequence; 
QPSK: quadrature phase-shift keying; IFFT: inverse fast Fourier transform; 
FFT: fast Fourier transform; FDE: frequency-domain equalization.}
\label{fig:dsp-flow}
\end{figure}

\section{Numerical Simulation}
\label{sec:numerical-simulation}
Numerical simulations first examine the signal-model predictions in Section~\ref{sec:principle}
before the experimental test. They isolate the mechanism and boundary of the proposed repeated
No-CP tail-reuse reconstruction. The simulations first test the range
condition: whether the repeated useful OFDM period provides the virtual cyclic extension required
for channel reconstruction when $N\ge M$, and how the reconstruction fails when this condition is
violated. Dynamic perturbation recovery then examines whether the same reconstruction chain can
support DAS vibration readout under the modeled conditions. As a secondary design study, bandwidth
scaling is examined at a fixed reporting interval.

The simulated transmitter and receiver digital signal processing (DSP) chain is summarized in
Fig.~\ref{fig:dsp-flow}. The transmitter maps a known pseudo-random binary sequence to quadrature
phase-shift keying (QPSK) symbols, generates the OFDM useful period, and repeats it without inserting
explicit CP samples. The receiver samples the backscattered signal, forms the virtual-CP observation
described in Section~\ref{subsec:periodic-nocp}, estimates the channel in the frequency domain,
transforms it back to range, and obtains the DAS phase map through gauge-differential processing.

\subsection{Simulation Configuration}
\label{subsec:simulation-configuration}
The simulated link contains 5.2~km of sensing fiber, corresponding to an approximately
52-$\mu$s round-trip channel memory. The reference OFDM bandwidth is set to 75~MHz, the
complex-baseband sampling rate is 75~MSa/s, and the useful-block length is 4096 samples.
This combination gives a 54.61-$\mu$s useful period, a 5.46-km unambiguous reconstruction span,
and a 1.333-m range-bin spacing. The corresponding channel memory is 3900 samples, so the useful
period slightly exceeds the physical fiber support. The same known OFDM useful block is transmitted
periodically without an explicit CP, cyclic suffix, or zero guard, and the preceding-period tail
provides the virtual prefix used by the receiver.

Unless otherwise stated, the tail-reuse, range-condition, and dynamic simulations use this 75-MHz
configuration. The bandwidth-scaling study in
Section~\ref{subsec:simulation-bandwidth-redundancy} instead varies the occupied bandwidth while
scaling the useful-block length proportionally.

The fiber backscatter is represented by a dense complex Rayleigh channel whose slow-time phase is
locally perturbed by prescribed vibration events. The channel is held constant within each OFDM
period and updated at the slow-time sampling instants. Additive receiver noise is included before
frequency-domain channel estimation. After FFT-domain deconvolution and inverse FFT reconstruction,
the recovered range-domain channel is converted into vibration traces through gauge-differential
phase processing. Table~\ref{tab:simulation-parameters} summarizes the parameters used for the
tail-reuse, range-condition, and dynamic simulations.

The nominal and shortened-period configurations test the valid and folded range conditions,
respectively. In the nominal case, $N=4096>M=3900$; the shortened-period case retains the same
reconstruction chain while deliberately violating this condition.

The continuous CP-OFDM reference used in the dynamic simulations transmits explicit-CP OFDM blocks
back to back. It uses the same 4096-sample useful block, prepends a 3899-sample CP, and removes the
CP before FFT-domain channel estimation. The otherwise matched reconstruction conditions isolate
the change in slow-time probing period introduced by the explicit CP.

\begin{table}[!t]
\caption{Parameters for the Tail-Reuse and Dynamic DAS Simulations}
\label{tab:simulation-parameters}
\centering
\begin{tabular}{@{}p{0.45\columnwidth}p{0.47\columnwidth}@{}}
\toprule
Parameter & Value or setting \\
\midrule
Fiber length & 5.2~km \\
Reference OFDM bandwidth & 75~MHz \\
Sampling rate & 75~MSa/s \\
Range-bin spacing & 1.333~m \\
Effective channel memory & 3900 samples \\
Gauge length & 2.667~m \\
OFDM useful length & 4096 samples \\
Unambiguous reconstruction range & 5.46~km \\
No-CP explicit CP length & 0 samples \\
Virtual prefix length & 3899 samples \\
No-CP useful-period duration & 54.61~$\mu$s \\
No-CP slow-time frames & 150 \\
No-CP Nyquist limit & 9.16~kHz \\
Receiver SNR & 45~dB \\
CP-OFDM explicit CP length & 3899 samples \\
CP-OFDM block period & 106.60~$\mu$s \\
CP-OFDM frames & 77 over the same 8.137-ms window \\
CP-OFDM Nyquist limit & 4.69~kHz \\
\bottomrule
\end{tabular}
\end{table}

\subsection{Tail-Reuse and Range-Condition Verification}
\label{subsec:simulation-tail-reuse-range}
\begin{figure}[!t]
\centering
\includegraphics[width=\columnwidth]{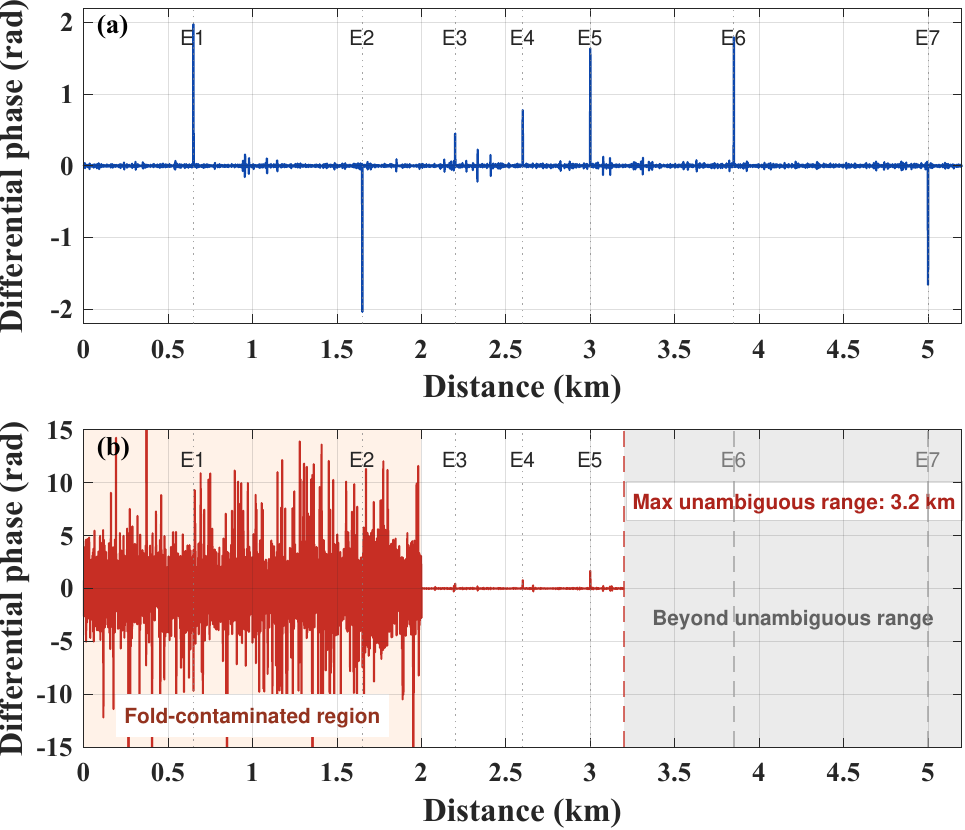}
\caption{Numerical comparison of valid repeated No-CP tail-reuse reconstruction and invalid $N<M$
range folding. (a) When the useful period covers the fiber memory, events E1--E7 remain localized
over the 5.2-km span. (b) With a 3.2-km unambiguous range, the response from the 3.2--5.2-km fiber
section folds into the reconstructed 0--2.0-km interval, while output beyond 3.2~km has no unique
physical-range interpretation. Both panels show the spatial differential-phase snapshot of the
109th repeated OFDM block.}
\label{fig:range-condition-simulation}
\end{figure}

The first simulation directly tests the range condition behind the virtual-CP reconstruction.
Seven perturbation events are placed along the
5.2-km fiber so that the reconstruction must assign responses over nearly the entire sensing span.
The event positions are 0.65, 1.65, 2.20, 2.60, 3.00, 3.85, and 5.00~km, and
the corresponding vibration frequencies are 500, 650, 800, 950, 1100, 1250, and
1400~Hz, with the same 1-rad phase-modulation amplitude. The plotted traces in
Fig.~\ref{fig:range-condition-simulation} are single-block spatial snapshots selected from the
109th repeated OFDM block, i.e., the 109th slow-time frame of a 150-frame simulation sequence.
In the valid configuration, $N=4096$ and $M=3900$, so the useful OFDM period is long enough to cover
the effective fiber memory. The tail of the preceding useful period can therefore supply the cyclic
extension required by the current receive window, and the expected outcome is a spatial response
localized at the physical event positions.

Figure~\ref{fig:range-condition-simulation}(a) follows this prediction. The labeled events
E1--E7 are recovered at their prescribed distances, while the inter-event regions remain close to
the background level. This behavior is the numerical counterpart of
\eqref{eq:circular-channel}: after virtual-CP construction, the receiver observes an effective
circular convolution over the physical fiber support and can reconstruct the range-domain channel by
FFT-domain deconvolution and inverse FFT processing.

The same figure also shows the failure mode when the condition $N\ge M$ is deliberately violated.
In this invalid case, a short-period branch with $N_{\mathrm{fold}}<M$ is used, giving a 3.2-km
maximum unambiguous range that is smaller than the 5.2-km fiber length. As predicted by
\eqref{eq:folded-channel}, the complex channel response from the physical 3.2--5.2-km section folds
into the reconstructed 0--2.0-km interval. Gauge-differential phase is calculated after this
complex-channel superposition, so the resulting cross terms distort the phase response within the
preceding range interval instead of producing simply shifted event replicas. Output beyond 3.2~km
lies outside the valid reconstruction range and therefore has no unique physical-range interpretation.

\begin{figure}[!t]
\centering
\includegraphics[width=0.9\columnwidth]{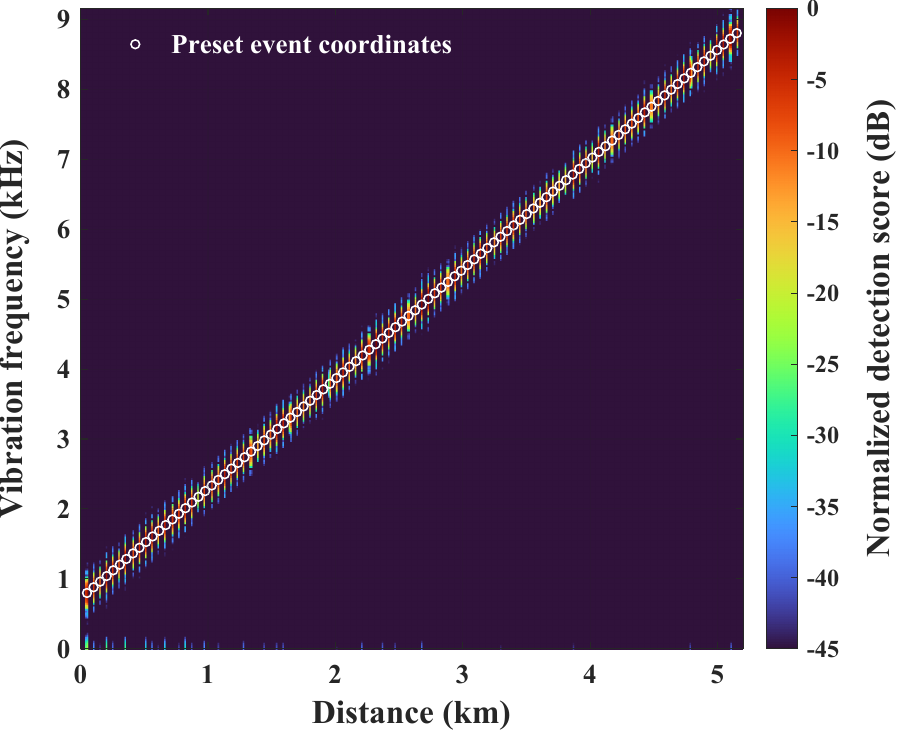}
\caption{Distance--frequency detection map for the 100-event repeated No-CP OFDM DAS simulation.
The score combines median-normalized slow-time spectral magnitude with square-root-weighted spatial
root-mean-square phase activity. White circles mark the prescribed event coordinates from 50~m to
5150~m and from 800~Hz to 8800~Hz.}
\label{fig:dynamic-multifrequency-simulation}
\end{figure}

\subsection{Dynamic Sensing and Slow-Time Bandwidth}
\label{subsec:simulation-dynamic-bandwidth}
After the range condition has been tested, the same reconstruction chain is applied to dynamic
perturbation scenes. A 100-event scene is used here to check whether the tail-reuse receiver can
recover a dense set of simultaneous vibration responses over the full sensing span. The vibration
positions are uniformly distributed from 50~m to 5150~m, and the vibration frequencies are
uniformly swept from 800~Hz to 8800~Hz. All events have the same 0.5-rad phase-modulation
amplitude with random initial phases. The tail-reuse No-CP reconstruction uses 150 slow-time frames
and gives an 8.137-ms monitoring window. This frequency range remains below the 9.16-kHz No-CP slow-time
Nyquist limit. For comparison, the continuous CP-OFDM reference follows the schedule defined in
Section~\ref{subsec:simulation-configuration}. It provides 77 slow-time samples over the same
monitoring window, with a 106.60-$\mu$s block period and a 4.69-kHz Nyquist limit.

Figure~\ref{fig:dynamic-multifrequency-simulation} presents the reconstructed scene in the
distance--frequency plane. At each distance cell, the slow-time spectral magnitude is normalized by
its non-DC median and weighted by the square root of the normalized spatial root-mean-square phase
activity. The resulting score is normalized to its global maximum for visualization. The bright
narrowband responses follow the 100 prescribed distance--frequency coordinates across the full
5.2-km span. Their alignment with the preset markers shows that the tail-reuse receiver preserves
the joint range and frequency assignment of the simulated events, consistent with the valid range
condition in Fig.~\ref{fig:range-condition-simulation}.

\begin{figure}[!t]
\centering
\includegraphics[width=\columnwidth]{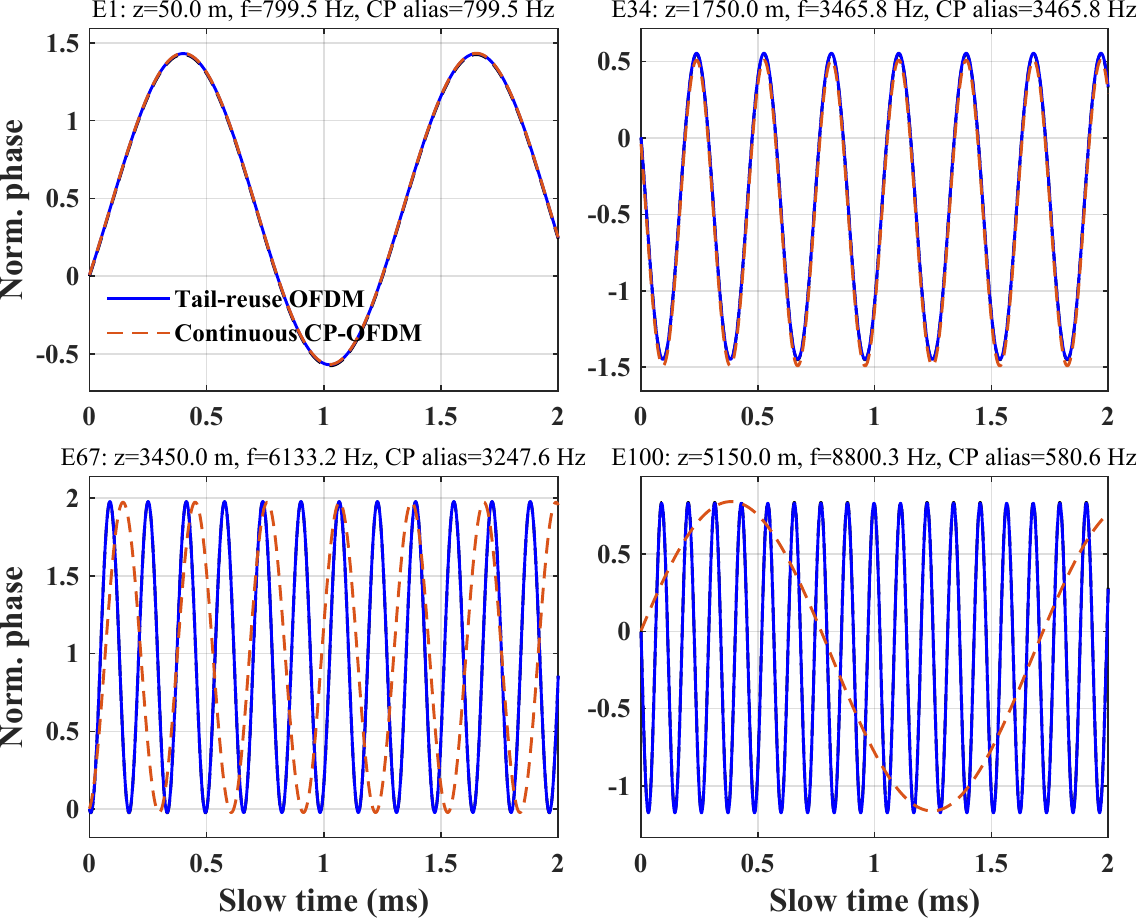}
\caption{Recovered vibration waveforms for representative events in the 100-event simulation.
The No-CP tail-reuse result is compared with the continuous CP-OFDM reference over a 2-ms
slow-time window. The panel titles list the event position, prescribed vibration frequency, and
continuous CP-OFDM alias frequency.}
\label{fig:waveform-comparison-simulation}
\end{figure}

The joint distance--frequency result in Fig.~\ref{fig:dynamic-multifrequency-simulation} is then
examined in the slow-time domain through representative traces.
Figure~\ref{fig:waveform-comparison-simulation} compares
four events selected from the same 100-event simulation. For the events near 0.8~kHz and 3.5~kHz,
both the No-CP tail-reuse receiver and the continuous CP-OFDM reference remain below the CP-OFDM
Nyquist limit, so the displayed waveforms are close. For the events near 6.1~kHz and 8.8~kHz,
however, the continuous CP-OFDM slow-time sampling rate is below the requirement and the displayed
waveforms appear at the corresponding aliased frequencies. The No-CP traces continue to follow the
high-frequency oscillations because the useful-block repetition period provides the larger
slow-time sampling rate.

\begin{figure}[!t]
\centering
\includegraphics[width=0.7\columnwidth]{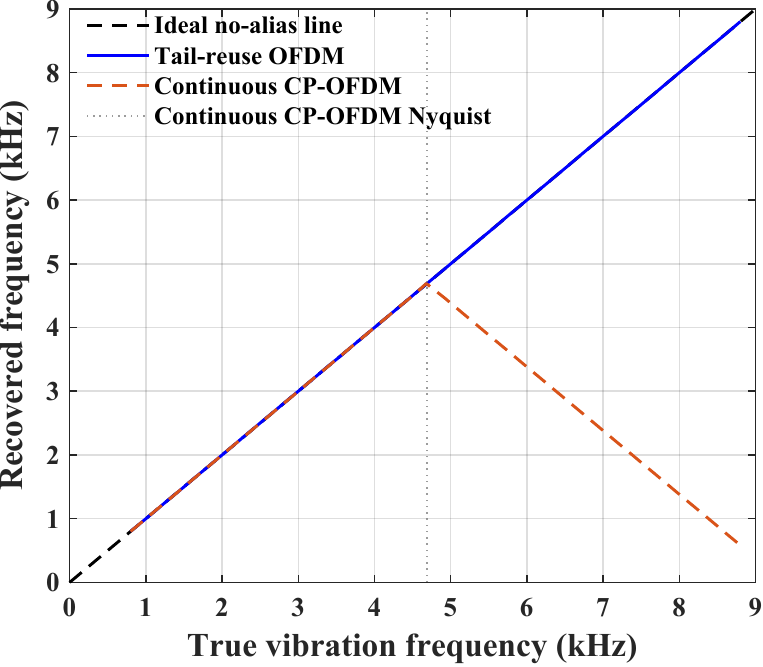}
\caption{Recovered vibration frequency versus prescribed vibration frequency for 50 representative
events selected from the 100-event simulation. The dotted vertical line marks the continuous
CP-OFDM Nyquist limit.}
\label{fig:frequency-recovery-simulation}
\end{figure}

The same trend is summarized in the frequency domain in Fig.~\ref{fig:frequency-recovery-simulation}. 
The recovered dominant frequency is plotted against
the prescribed vibration frequency for 50 representative events selected from the 100-event scene.
The No-CP tail-reuse result follows the ideal no-alias relation across the 800-Hz to 8800-Hz
frequency span. By contrast, the continuous CP-OFDM result follows the ideal line only below its
4.69-kHz Nyquist limit and then folds to lower recovered frequencies. This frequency-domain view is
consistent with the waveform examples in Fig.~\ref{fig:waveform-comparison-simulation} and shows
that the high-frequency recovery is a consequence of the larger slow-time sampling rate provided by
the repeated useful OFDM period.

\subsection{Bandwidth Scaling at a Fixed Reporting Interval}
\label{subsec:simulation-bandwidth-redundancy}

For this design study, the occupied bandwidth is increased from 75~MHz to 150, 225, and 300~MHz,
with the receiver processing rate increased correspondingly from 75~MSa/s to 150, 225, and
300~MSa/s. The useful-block length is scaled proportionally from 4096 samples to 8192, 12288, and
16384 samples. The useful-period
duration, unambiguous sensing range, slow-time sampling rate, and final reporting interval therefore
remain fixed, while the reconstructed range response is sampled on progressively finer spatial
grids. Locally consistent fine-grid gauge-phase observations are then combined within each reporting
cell without temporal averaging.

\begin{figure}[!t]
\centering
\includegraphics[width=\columnwidth]{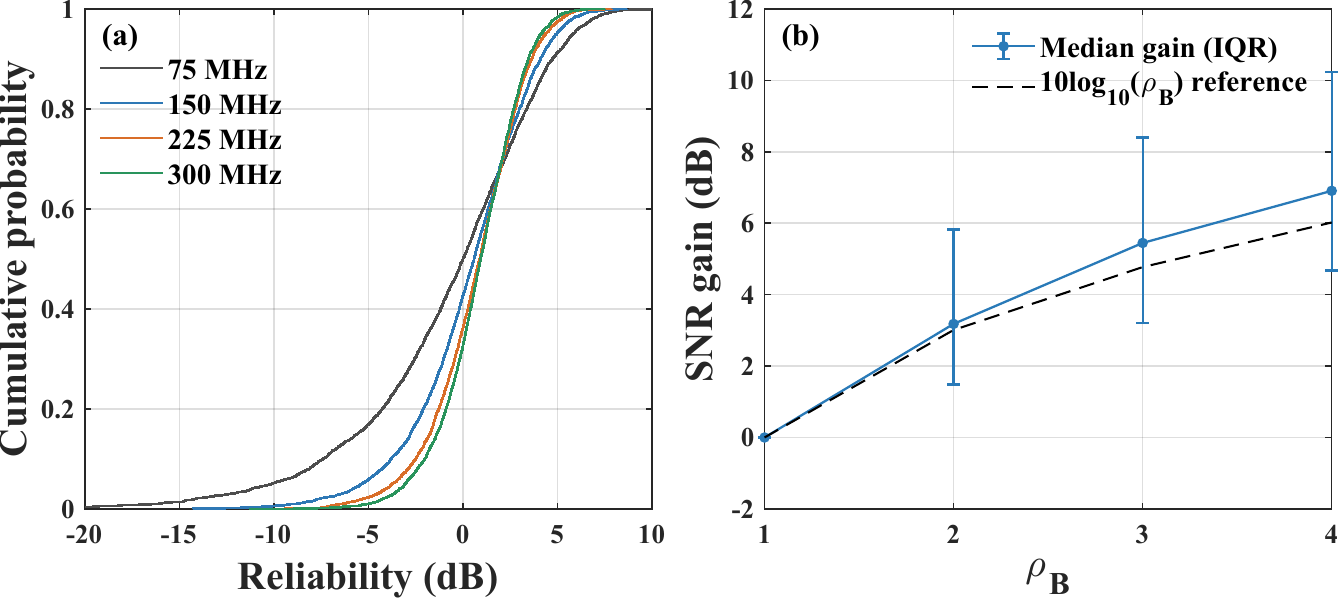}
\caption{Bandwidth scaling at a fixed reporting interval in repeated No-CP OFDM DAS simulation. (a) Effective
differential-phase reliability distributions for 75--300~MHz after joint processing to a common
reporting interval, normalized according to \eqref{eq:effective-differential-phase-reliability}.
(b) Within-band reconstruction-SNR gain of spatial joint processing over a single fine-grid trace
versus $\rho_B$. Markers and error bars denote the event median and interquartile range; the dashed
line is the nominal $10\log_{10}(\rho_B)$ equal-reliability reference.}
\label{fig:bandwidth-redundancy-simulation}
\end{figure}

For the reliability distribution in Fig.~\ref{fig:bandwidth-redundancy-simulation}(a), an
effective differential-phase reliability is used as a mechanism-level diagnostic. Let $h_i$ and
$h_{i+G}$ denote two complex Rayleigh channel samples separated by the gauge spacing $G$. The
single fine-sample reliability, the joint reliability of reporting cell $\mathcal{C}_j$, and the
normalized dB quantity plotted in Fig.~\ref{fig:bandwidth-redundancy-simulation}(a) are defined as
\begin{equation}
\begin{gathered}
r_i = \frac{|h_i|^2 |h_{i+G}|^2}
{|h_i|^2+|h_{i+G}|^2+\epsilon},\\[10pt]
R_j^{(\rho_B)} = \sum_{i\in\mathcal{C}_j} r_i,\\
\eta_j^{(\rho_B)} =
10\log_{10}\frac{R_j^{(\rho_B)}}{\operatorname{median}_j R_j^{(1)}} .
\end{gathered}
\label{eq:effective-differential-phase-reliability}
\end{equation}
This metric reflects the fact that a gauge-differential phase estimate becomes unstable when either
end of the gauge pair falls into a Rayleigh fading null. Accordingly, the metric characterizes the
fading-robustness mechanism of spatial joint processing.

Figure~\ref{fig:bandwidth-redundancy-simulation} evaluates this spatial-combining mechanism. In
Fig.~\ref{fig:bandwidth-redundancy-simulation}(a), increasing the occupied bandwidth from 75~MHz to
300~MHz shifts the reliability distribution to higher values while reducing the low-reliability
tail. This indicates that combining more fine spatial observations lowers the probability that a
reported gauge-differential phase sample is dominated by a local Rayleigh fading null.
Figure~\ref{fig:bandwidth-redundancy-simulation}(b) gives the within-band reconstruction-SNR gain
from spatial joint processing relative to a single fine-grid trace at the same occupied bandwidth.
The median gain increases with $\rho_B$ and follows the same trend as the nominal
equal-reliability averaging reference, although the wide interquartile ranges show that the gain is
scene- and fading-dependent. Deviations from the dashed line are consistent with nonuniform Rayleigh
reliability and the weighted spatial combining used in the reconstruction. These results indicate
that additional spatial samples can mitigate fading-induced weak points. Because the baseline in
Fig.~\ref{fig:bandwidth-redundancy-simulation}(b) is a single fine-grid trace at the
same occupied bandwidth, the comparison isolates the spatial-combining gain from the bandwidth
change. The result therefore characterizes spatial combining at a fixed reporting resolution under
the modeled noise and fading conditions.

These numerical results are consistent with the design consequence in
Section~\ref{subsec:design-consequences}. Reducing the transmitted period from an explicit-CP frame
to the OFDM useful period increases the slow-time sampling rate for the same useful-block length
and fiber memory. In the present 5.2-km simulation configuration, the No-CP period gives a 9.16-kHz
slow-time Nyquist limit. The comparison therefore supports a period-limited sensing-bandwidth benefit,
and quantifies the effect of removing the explicitly transmitted CP from the probing period.

\section{Experimental Setup and Verification}
\label{sec:experimental-setup}
\subsection{Experimental Setup}
The experiment evaluates the repeated No-CP OFDM DAS waveform and tail-reuse receiver over a 5.2-km
fiber link. The optical architecture follows Fig.~\ref{fig:system-schematic} and is similar to the
heterodyne coherent platform used in the precursor CP-OFDM experiment \cite{huang2026cpofdmisi},
with a continuously repeated No-CP OFDM useful block used as the probe.

\begin{table}[!t]
\caption{Experimental Parameters of the Repeated No-CP OFDM DAS System}
\label{tab:parameters}
\centering
\begin{tabular}{p{0.43\columnwidth}p{0.47\columnwidth}}
\toprule
Parameter & Value \\
\midrule
Laser wavelength & 1550~nm \\
Laser linewidth & 1~kHz \\
AWG sampling rate & 512~MSa/s \\
Oscilloscope sampling rate & 1~GSa/s \\
Fiber length & 5.2~km \\
Baseband processing rate & 128~MSa/s \\
Digital IQ shift & 160~MHz \\
Occupied OFDM bandwidth & 111.984~MHz \\
Active/total subcarriers & 7167/8192 \\
Subcarrier spacing & 15.625~kHz \\
Edge guard bandwidth per side & 8~MHz \\
OFDM size & 8192 samples \\
CP length & 0 samples \\
OFDM useful-block period & 64.0~$\mu$s \\
No-CP slow-time Nyquist limit & 7.81~kHz \\
RF bandpass & 94--226~MHz \\
Nominal gauge length & 4~m \\
RVS diversity channels & 33 \\
Blind search band & 300--7800~Hz \\
\bottomrule
\end{tabular}
\end{table}

A narrow-linewidth laser at 1550~nm provides the optical carrier. In the signal path, a Keysight
M8190A arbitrary waveform generator (AWG) drives an ID Photonics OMFT-40 in-phase/quadrature (IQ)
modulator to generate the repeated No-CP OFDM probe. The probe is amplified by an erbium-doped fiber
amplifier (EDFA) and launched into the sensing fiber with an input optical power of -11~dBm. The
Rayleigh backscatter returns through the circulator path, is detected by a 300-MHz balanced
photodetector (BPD), and is digitized by a RIGOL DHO4404 oscilloscope for offline processing. The
numerical waveform, link, and acquisition parameters are summarized in
Table~\ref{tab:parameters}.

The digital probe is generated on a 128-MSa/s complex-baseband grid with an 8192-sample useful OFDM
block. The repeated No-CP OFDM waveform does not activate all FFT bins. An 8-MHz guard band is reserved at
each spectral edge, and the DC subcarrier is nulled. As a result, 7167 of the 8192 subcarriers are
active, giving an occupied OFDM bandwidth of 111.984~MHz. The experimental range response therefore
follows the band-limited kernel in \eqref{eq:effective-reconstruction-kernel}. The baseband waveform
is shifted to a 160-MHz intermediate frequency and exported to the AWG at 512~MSa/s. Each AWG
segment contains three identical useful periods and is looped continuously. The 64.0-$\mu$s useful
period gives a 15.625-kHz slow-time sampling rate and a 7.81-kHz Nyquist limit.

The vibration is applied by a piezoelectric transducer (PZT) inserted after approximately 5.1~km of
fiber and followed by an additional 100-m fiber segment. This
layout-defined position is treated as nominal because no independent absolute-distance calibration
was performed. The following results use two representative 5-V sinusoidal PZT records, at 500~Hz
and 3~kHz, acquired at the same perturbation position.

\begin{figure}[!t]
\centering
\includegraphics[width=0.9\columnwidth]{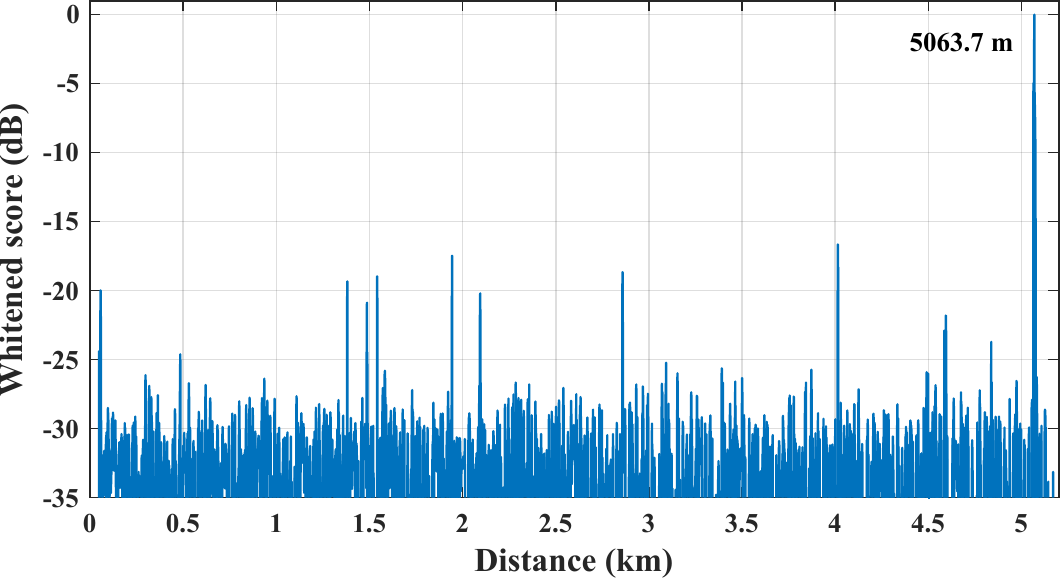}\\[0.3cm]
\includegraphics[width=0.9\columnwidth]{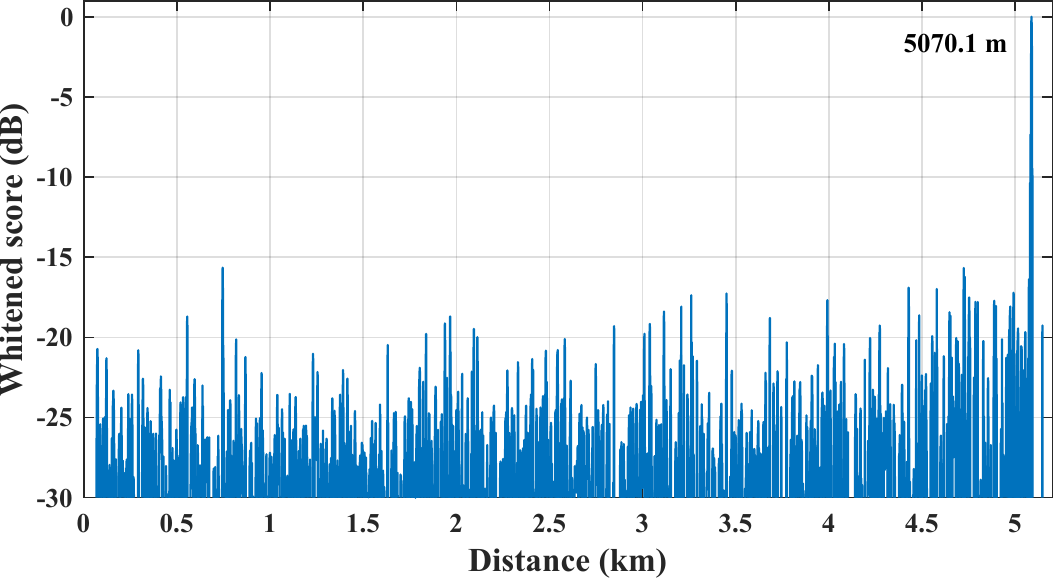}
\caption{Blind vibration localization in the 5.2-km repeated No-CP OFDM DAS experiment. Top: 500-Hz
PZT record with a dominant spatially whitened score peak at 5063.7~m. Bottom: 3-kHz PZT record with
a dominant peak at 5070.1~m.}
\label{fig:experimental-localization}
\end{figure}

\begin{figure*}[!t]
\centering
\begin{minipage}[t]{0.49\textwidth}
\centering
\includegraphics[width=0.95\linewidth]{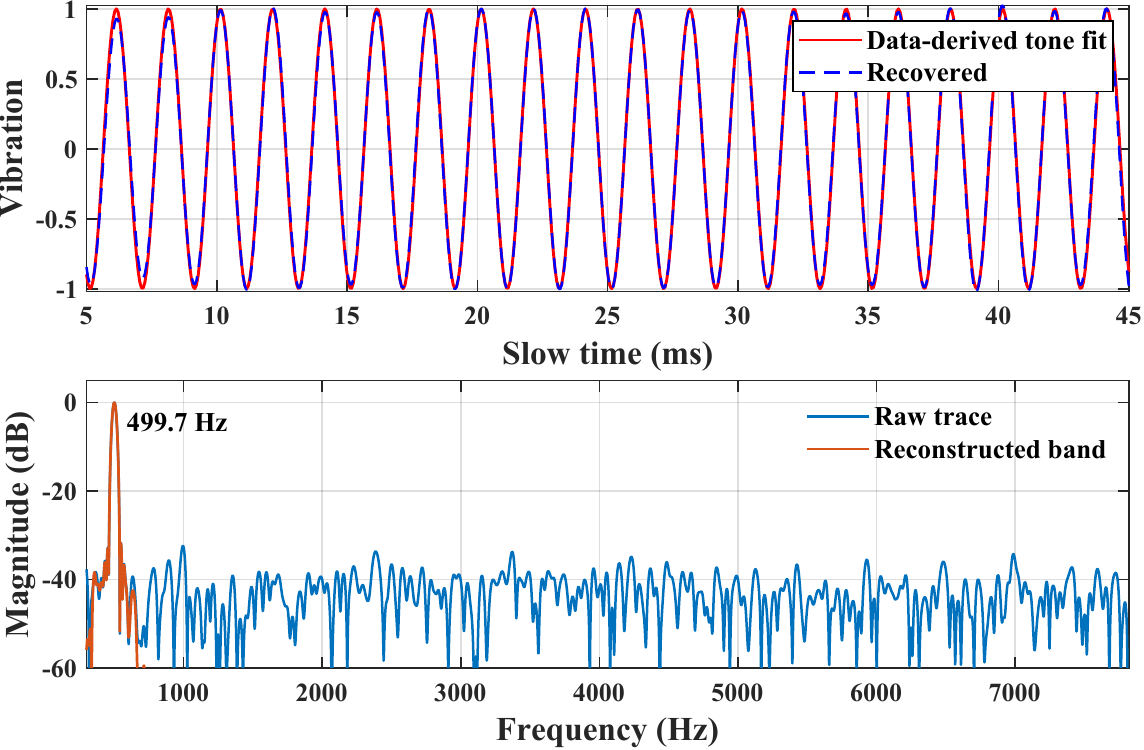}\\[0.3cm]
\includegraphics[width=0.95\linewidth]{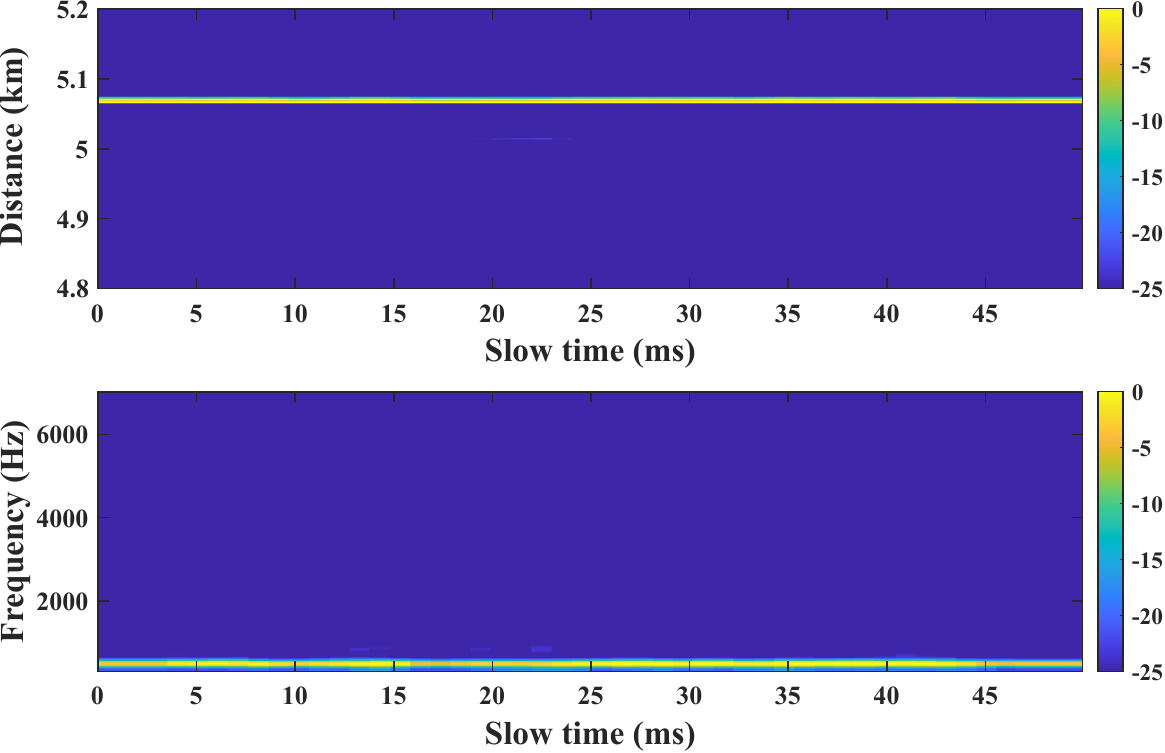}
\caption{Experimental recovery of the 500-Hz PZT vibration. Upper composite: normalized recovered 
slow-time waveform at the detected range bin, data-derived tone fit, and vibration spectrum with a
dominant peak at 499.7~Hz. Lower composite: full-record distance-time and frequency-time maps,
showing that the recovered response remains near the detected PZT region and near the 500-Hz drive
over the acquisition interval.}
\label{fig:experimental-500hz-recovery}
\end{minipage}
\hfill
\begin{minipage}[t]{0.49\textwidth}
\centering
\includegraphics[width=0.95\linewidth]{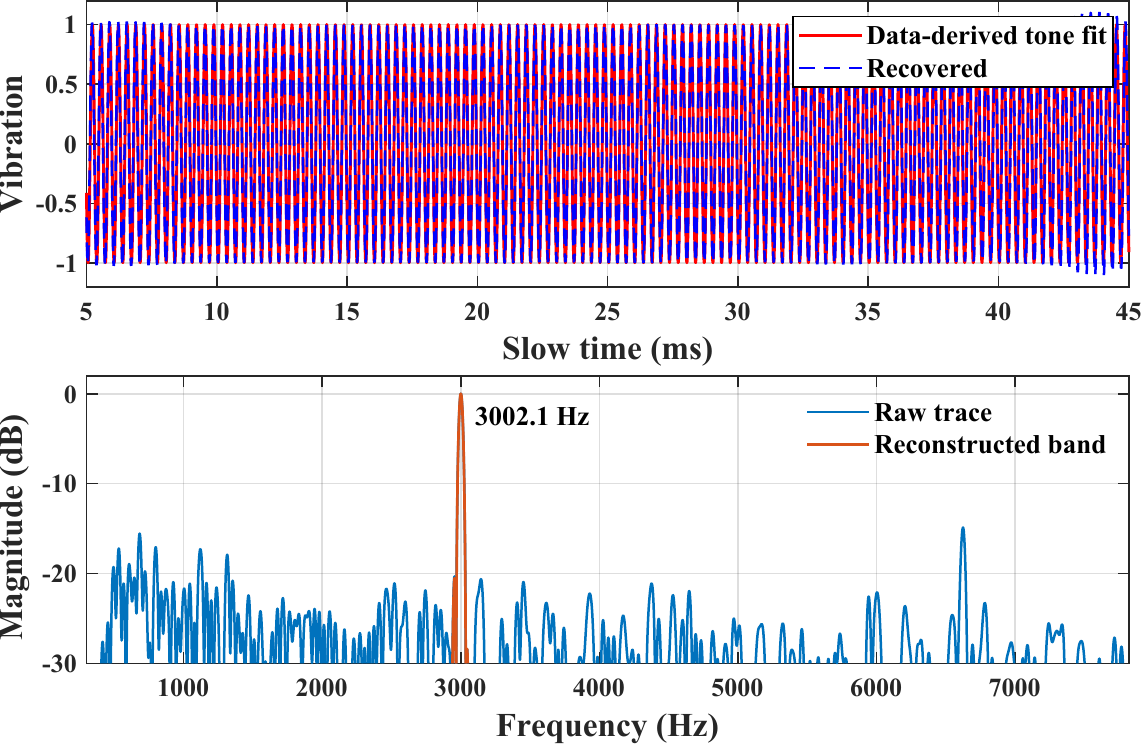}\\[0.3cm]
\includegraphics[width=0.95\linewidth]{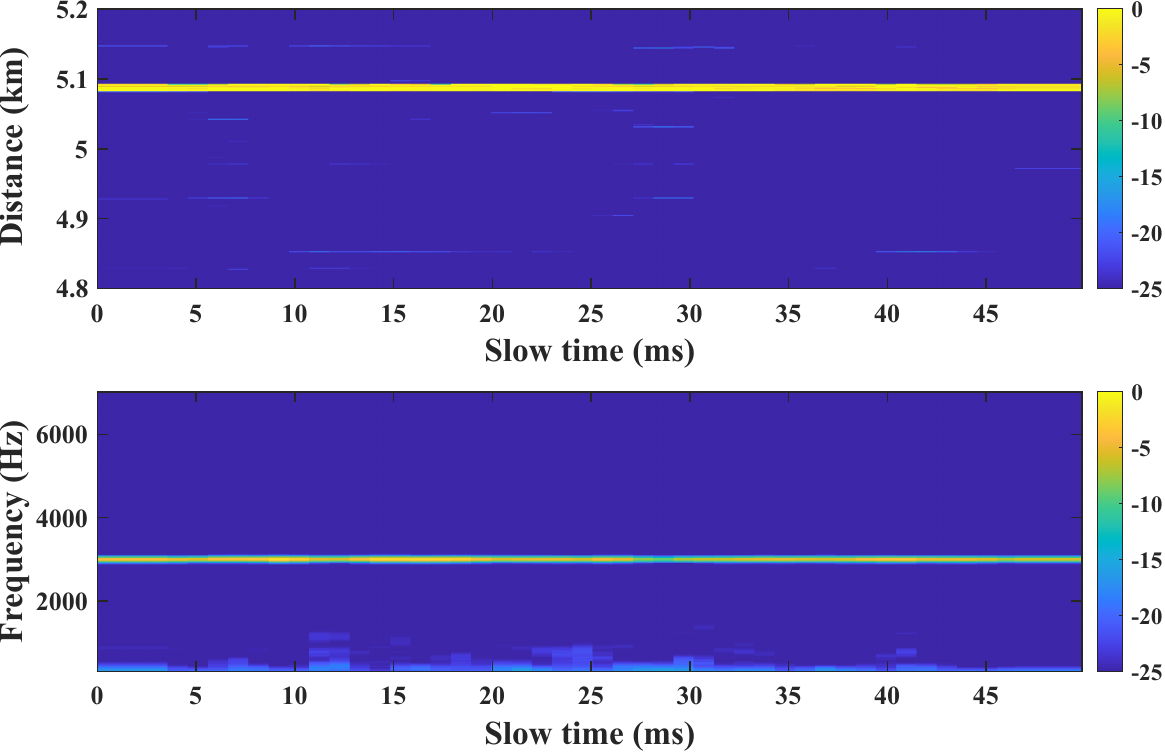}
\caption{Experimental recovery of the 3-kHz PZT vibration. Upper composite: normalized recovered
slow-time waveform at the detected range bin, data-derived tone fit, and vibration spectrum with a
dominant component near the applied 3-kHz drive. Lower composite: full-record distance-time and
frequency-time maps, showing that the recovered response remains near the detected PZT region and
near 3~kHz over the acquisition interval.}
\label{fig:experimental-3000hz-recovery}
\end{minipage}
\end{figure*}

The acquired oscilloscope waveform is processed offline using the DSP chain in
Fig.~\ref{fig:dsp-flow} and the exact repeated No-CP OFDM reference saved by the transmitter. The waveform is
first mean-removed, normalized, filtered by a 94--226-MHz radio-frequency (RF) bandpass, digitally
downconverted at 160~MHz, and resampled to the 128-MSa/s OFDM processing grid. Because the received
waveform contains neither an explicit CP nor a zero guard, the period boundary is not obtained from
a CP correlation peak. Instead, candidate window starts are evaluated using the support-to-tail
energy criterion in \eqref{eq:tail-metric}. After frequency-domain deconvolution and inverse FFT
reconstruction, the selected boundary is the one that concentrates the reconstructed channel inside
the physical 0--5.2-km fiber support rather than in the circular FFT tail.

The selected one-period window is then processed by regularized active-subcarrier division and an
inverse FFT. Rayleigh fading is mitigated by frequency-subband rotated-vector summation (RVS) using
32 overlapping OFDM subbands together with the full-band estimate \cite{huang2026cpofdmisi}. The
reconstructed channel is converted to gauge-differential phase with a 4-m nominal gauge length, and
blind localization is performed from the spatially whitened slow-time spectrum over 300--7800~Hz.

\subsection{Experimental Verification}
\label{subsec:experimental-verification}

The driven fiber section is first localized without prior position or frequency information. For
each valid gauge position, the recovered slow-time gauge-differential phase is transformed into the
frequency domain over the search band
$\mathcal{B}=[300,7800]$~Hz. Let $P_\ell(f)$ denote the resulting slow-time power spectrum at
range bin $\ell$. To suppress the range-dependent background and fading-induced contrast
variations, the spectrum is spatially whitened at each frequency as
\begin{equation}
\widetilde{P}_{\ell}(f)=
\frac{P_{\ell}(f)}
{\mathrm{median}_{m\in\mathcal{V}}\{P_m(f)\}+\epsilon},
\label{eq:experimental-whitening}
\end{equation}
where $\mathcal{V}$ is the set of valid range bins and $\epsilon$ is a small regularization term.
The blind localization score plotted in Fig.~\ref{fig:experimental-localization} is then defined as
\begin{equation}
S_{\ell}^{\mathrm{dB}}=
10\log_{10}
\frac{\max_{f\in\mathcal{B}}\widetilde{P}_{\ell}(f)}
{\max_{m\in\mathcal{V}}\max_{f\in\mathcal{B}}\widetilde{P}_{m}(f)} .
\label{eq:experimental-whitened-score}
\end{equation}
The frequency-wise median provides a background reference across the valid fiber span, and the
maximization over $\mathcal{B}$ selects the strongest whitened narrowband response at each position.
The score is therefore obtained solely from the measured broadband response, without inserting the
nominal PZT position or the applied drive tone. The same procedure is applied independently to the
500-Hz and 3-kHz records.

As shown in Fig.~\ref{fig:experimental-localization}, the 500-Hz record gives a dominant score peak
at 5063.7~m, and the 3-kHz record gives a dominant score peak at 5070.1~m. The two independently
obtained positions differ by 6.4~m and both fall near the nominal PZT region at the distal end of the
5.2-km link. Because the PZT position is defined by the fiber layout without an independent
absolute-distance calibration, this separation serves as a measure of inter-record localization
consistency; no absolute localization error is assigned. The two distal responses therefore support
a common physical origin within the driven fiber section.

After the perturbation position is obtained, the slow-time gauge-differential phase is extracted at
the detected range bin for temporal, spectral, and full-record analysis. The two records are
processed independently at their respective detected bins, with no further spatial search during
waveform or time-frequency recovery. Each frequency case is therefore evaluated using the range
location returned by its own blind localization result.

% \begin{figure}[!t]
% \centering
% \includegraphics[width=0.9\columnwidth]{figures/Fig19_500.pdf}
% \vspace{5mm}
% \includegraphics[width=0.9\columnwidth]{figures/Fig17_500.pdf}
% \caption{Experimental recovery of the 500-Hz PZT vibration. Upper composite: normalized recovered
% slow-time waveform at the detected range bin, data-derived tone fit, and vibration spectrum with a
% dominant peak at 499.7~Hz. Lower composite: full-record distance-time and frequency-time maps,
% showing that the recovered response remains near the detected PZT region and near the 500-Hz drive
% over the acquisition interval.}
% \label{fig:experimental-500hz-recovery}
% \end{figure}

Figure~\ref{fig:experimental-500hz-recovery} presents the recovery result for the 500-Hz
record. At the detected range bin, the recovered trace follows the data-derived sinusoidal
component, and the spectrum has a distinct maximum at 499.7~Hz, 0.3~Hz from the nominal drive. The
agreement in the time and frequency domains shows that the recovered phase evolution retains the
applied low-frequency oscillation. Over the full acquisition interval, the distance-time response
remains near the localized distal region, while the frequency-time response remains concentrated
around 500~Hz. These two maps extend the single-trace result to the complete record and show that the
recovered component remains spatially localized and temporally persistent.

% \begin{figure}[!h]
% \centering
% \includegraphics[width=0.9\columnwidth]{figures/Fig19_3000.pdf}
% \vspace{1mm}
% \includegraphics[width=0.9\columnwidth]{figures/Fig17_3000.pdf}
% \caption{Experimental recovery of the 3-kHz PZT vibration. Upper composite: normalized recovered
% slow-time waveform at the detected range bin, data-derived tone fit, and vibration spectrum with a
% dominant component near the applied 3-kHz drive. Lower composite: full-record distance-time and
% frequency-time maps, showing that the recovered response remains near the detected PZT region and
% near 3~kHz over the acquisition interval.}
% \label{fig:experimental-3000hz-recovery}
% \end{figure}

Figure~\ref{fig:experimental-3000hz-recovery} provides the corresponding recovery evidence for the
3-kHz record. The recovered waveform contains the expected denser oscillations, and its spectrum is
concentrated at 3002.1~Hz, within the 7.81-kHz slow-time Nyquist limit. The close agreement with the
applied drive confirms that the 3-kHz phase evolution is retained by the tail-reuse
receiver. The full-record distance-time map remains concentrated near the nominal PZT region, and
the frequency-time map forms a persistent ridge near 3~kHz over successive slow-time windows.

The two records also impose distinct slow-time sampling demands. At the 15.625-kHz slow-time
sampling rate, the 500-Hz and 3-kHz vibrations are represented by approximately 31 and 5 samples per
cycle, respectively. The recovered peaks at 499.7 and 3002.1~Hz correspond to relative deviations
of approximately 0.06\% and 0.07\% from the applied frequencies. The 3-kHz case therefore evaluates
phase recovery with substantially fewer slow-time observations per vibration cycle while remaining
below the 7.81-kHz Nyquist limit.

Together, the independently processed records show that the same tail-reuse receiver preserves
blind localization, temporal phase evolution, and frequency recovery under different slow-time
sampling demands. The persistent distance-time and frequency-time responses further confirm that
the recovered vibrations remain associated with the same physical fiber section throughout the
acquisition interval.

\section{Conclusion}
\label{sec:conclusion}
This paper proposed and analyzed a repeated cyclic-prefix-free OFDM (No-CP OFDM) DAS waveform and
tail-reuse receiver for coherent distributed vibration sensing. Under the valid range condition,
the receiver obtains the required cyclic extension from the preceding repeated useful-period tail
instead of transmitting a separate CP. The resulting useful-period FFT window supports
frequency-domain channel reconstruction and subsequent gauge-differential phase recovery. If the
useful period does not cover the physical channel memory, the reconstructed response becomes
circularly folded. The occupied bandwidth continues to determine the spatial resolution.

Removing the explicit CP shortens the slow-time probing period and raises the period-limited highest
unaliased vibration frequency. In the 5.2-km numerical configuration, the 54.61-$\mu$s No-CP period
gives a 9.16-kHz slow-time Nyquist limit, compared with 4.69~kHz for the sufficient-CP reference.
Simulations verified tail-reuse reconstruction, the range-folding boundary, and dynamic recovery in
a 100-event scene spanning 800~Hz to 8800~Hz. The No-CP receiver recovered the prescribed
frequencies across this range, whereas the continuous CP-OFDM reference produced aliased
frequencies above its lower Nyquist limit. This comparison demonstrates a period-limited
sensing-bandwidth benefit from eliminating the explicitly transmitted CP. A secondary
bandwidth-scaling simulation indicated that proportionally increasing the processing rate and
useful-block length can provide fine spatial observations for joint processing at a fixed reporting
interval under the modeled conditions.

Experiments applied the same receiver chain to a 5.2-km coherent DAS link. In
separate measurements with 5-V PZT excitation at 500~Hz and 3~kHz, the receiver blindly localized
the driven fiber region at 5063.7~m and 5070.1~m, respectively. Both positions are close to the
nominal PZT region near the distal end of the link. The recovered waveform and spectrum showed a
499.7-Hz peak for the 500-Hz record and a dominant component near the applied 3-kHz drive. These
experiments establish the feasibility of tail-reuse channel reconstruction and demonstrate
localization, waveform recovery, frequency recovery, and localized time-frequency tracking at both
drive frequencies. The analytical model and the matched numerical comparison with continuous
CP-OFDM support the corresponding extension of the period-limited unaliased vibration range.

% Add an acknowledgment and funding statement before submission, if applicable.

\bibliographystyle{IEEEtran}
\bibliography{references_v5}

\end{document}